\documentclass[twocolumn,nofootinbib,preprintnumbers,superscriptaddress,amsmath,amssymb,aps,prc,10pt]{revtex4-1}
\usepackage{dcolumn}
\usepackage{bm}
\usepackage[mathlines]{lineno}

\usepackage{multirow}
\usepackage{bm}
\usepackage{textcomp}
\usepackage{longtable}
\usepackage{booktabs}
\usepackage{epsfig}
\usepackage{rotating}
\usepackage{xfrac}
\usepackage{threeparttable}
\usepackage{lipsum} 
\usepackage{anyfontsize}
\usepackage{CJKutf8}

\usepackage{graphicx}
\usepackage{dcolumn}
\usepackage{float}
\usepackage{mathptmx, courier, pifont}
\usepackage[scaled=0.92]{helvet}
\usepackage[T1]{fontenc}
\usepackage{textcomp}
\usepackage{color}
\usepackage[normalem]{ulem}

\usepackage[dvipsnames]{xcolor}
\usepackage[colorlinks=true,urlcolor=Blue,linkcolor=Blue]{hyperref}
\usepackage[all]{hypcap}
\usepackage{makecell}

\usepackage{physics} 
\usepackage{comment} 

\begin{document}


\title{High-precision direct decay energy measurements of the electron-capture decay of $^{97}$Tc}
\author{Zhuang~Ge}
\thanks{Corresponding author}\email{zhuang.z.ge@jyu.fi}
\affiliation{Department of Physics, University of Jyv\"askyl\"a, P.O. Box 35, FI-40014, Jyv\"askyl\"a, Finland}%
\author{Tommi~Eronen}
\thanks{Corresponding author}\email{tommi.eronen@jyu.fi}
\affiliation{Department of Physics, University of Jyv\"askyl\"a, P.O. Box 35, FI-40014, Jyv\"askyl\"a, Finland}%
\author{Vasile~Alin~Sevestrean}
\thanks{Corresponding author}
 \email{sevestrean.alin@theory.nipne.ro}
\affiliation{International Centre for Advanced Training and Research in Physics, P.O. Box MG12, 077125 Bucharest-M\u{a}gurele, Romania}%
\affiliation{Faculty of Physics, University of Bucharest, 405 Atomiștilor, P.O. Box MG11, 077125 Bucharest-M\u{a}gurele, Romania}%
\affiliation{“Horia Hulubei” National Institute of Physics and Nuclear Engineering, 30 Reactorului, POB MG-6, RO-077125 Bucharest-M\u{a}gurele, Romania}
\author{Marlom~Ramalho}
\affiliation{Department of Physics, University of Jyv\"askyl\"a, P.O. Box 35, FI-40014, Jyv\"askyl\"a, Finland}%
\author{Ovidiu~Ni\c{t}escu}
\affiliation{International Centre for Advanced Training and Research in Physics, P.O. Box MG12, 077125 Bucharest-M\u{a}gurele, Romania}%
\affiliation{“Horia Hulubei” National Institute of Physics and Nuclear Engineering, 30 Reactorului, POB MG-6, RO-077125 Bucharest-M\u{a}gurele, Romania}
\author{Stefan Ghinescu}
\affiliation{International Centre for Advanced Training and Research in Physics, P.O. Box MG12, 077125 Bucharest-M\u{a}gurele, Romania}%
\affiliation{Faculty of Physics, University of Bucharest, 405 Atomiștilor, P.O. Box MG11, 077125 Bucharest-M\u{a}gurele, Romania}
\affiliation{“Horia Hulubei” National Institute of Physics and Nuclear Engineering, 30 Reactorului, POB MG-6, RO-077125 Bucharest-M\u{a}gurele, Romania}
\author{Sabin~Stoica}
\affiliation{International Centre for Advanced Training and Research in Physics, P.O. Box MG12, 077125 Bucharest-M\u{a}gurele, Romania}%
\author{Jouni~Suhonen}
\thanks{Corresponding author} %
\email{jouni.t.suhonen@jyu.fi}
\affiliation{Department of Physics, University of Jyv\"askyl\"a, P.O. Box 35, FI-40014, Jyv\"askyl\"a, Finland}%
\affiliation{International Centre for Advanced Training and Research in Physics, P.O. Box MG12, 077125 Bucharest-M\u{a}gurele, Romania}%
\author{Antoine~de Roubin}\thanks{Present address: Universit\'e de Caen Normandie, CNRS/IN2P3, LPC Caen UMR6534, F-14000 Caen, France}
\affiliation{KU Leuven, Instituut voor Kern- en Stralingsfysica, B-3001 Leuven, Belgium}%
\affiliation{Universit\'e de Bordeaux, CNRS/IN2P3, LP2I Bordeaux, UMR 5797, F-33170 Gradignan, France}%
\author{Dmitrii~Nesterenko}
\affiliation{Department of Physics, University of Jyv\"askyl\"a, P.O. Box 35, FI-40014, Jyv\"askyl\"a, Finland}%
\author{Anu~Kankainen}
\affiliation{Department of Physics, University of Jyv\"askyl\"a, P.O. Box 35, FI-40014, Jyv\"askyl\"a, Finland}%
\author{Pauline~Ascher}
\affiliation{Universit\'e de Bordeaux, CNRS/IN2P3, LP2I Bordeaux, UMR 5797, F-33170 Gradignan, France}%
\author{Samuel Ayet San~Andres}
\affiliation{Instituto de Fisica Corpuscular, CSIC-UV, 46980, Gradignan, Spain}
\author{Olga~Beliuskina}
\affiliation{Department of Physics, University of Jyv\"askyl\"a, P.O. Box 35, FI-40014, Jyv\"askyl\"a, Finland}%
\author{Pierre~Delahaye}
\affiliation{GANIL, CEA/DSM-CNRS/IN2P3, Bd Henri Becquerel, 14000 Caen, France}
\author{Mathieu~Flayol}
\affiliation{Universit\'e de Bordeaux, CNRS/IN2P3, LP2I Bordeaux, UMR 5797, F-33170 Gradignan, France}%
\author{Mathias~Gerbaux}
\affiliation{Universit\'e de Bordeaux, CNRS/IN2P3, LP2I Bordeaux, UMR 5797, F-33170 Gradignan, France}%
\author{St\'ephane~Gr\'evy}
\affiliation{Universit\'e de Bordeaux, CNRS/IN2P3, LP2I Bordeaux, UMR 5797, F-33170 Gradignan, France}%
\author{Marjut~Hukkanen}
\affiliation{Department of Physics, University of Jyv\"askyl\"a, P.O. Box 35, FI-40014, Jyv\"askyl\"a, Finland}%
\affiliation{Centre d'Etudes Nucl\'eaires de Bordeaux Gradignan, UMR 5797 CNRS/IN2P3 - Universit\'e de Bordeaux, 19 Chemin du Solarium, CS 10120, F-33175 Gradignan Cedex, France}
\author{Arthur~Jaries}
\affiliation{Department of Physics, University of Jyv\"askyl\"a, P.O. Box 35, FI-40014, Jyv\"askyl\"a, Finland}%
\author{Ari~Jokinen} 
\affiliation{Department of Physics, University of Jyv\"askyl\"a, P.O. Box 35, FI-40014, Jyv\"askyl\"a, Finland}%
\author{Audric~Husson} 
\affiliation{Universit\'e de Bordeaux, CNRS/IN2P3, LP2I Bordeaux, UMR 5797, F-33170 Gradignan, France}%
\author{Daid~Kahl}\thanks{Present address: Facility for Rare Isotope Beams, Michigan State University, 640 South Shaw Lane East Lansing, MI 48824, USA}
\affiliation{Extreme Light Infrastructure - Nuclear Physics, Horia Hulubei National Institute for R\&D in Physics and Nuclear Engineering (IFIN-HH), Bucharest-Magurele 077125, Romania}
\author{Joel~Kostensalo}
\affiliation{Natural Resources Institute Finland, Yliopistokatu 6B, FI-80100, Joensuu, Finland}%
\author{Jenni~Kotila}
\affiliation{Finnish Institute for Educational Research, University of Jyv\"askyl\"a, P.O. Box 35, FI-40014, Jyv\"askyl\"a, Finland}%
\affiliation{Center for Theoretical Physics, Sloane Physics Laboratory Yale University, New Haven, Connecticut 06520-8120, USA}
\author{Iain~Moore}
\affiliation{Department of Physics, University of Jyv\"askyl\"a, P.O. Box 35, FI-40014, Jyv\"askyl\"a, Finland}%
\author{Stylianos~Nikas}
\affiliation{Department of Physics, University of Jyv\"askyl\"a, P.O. Box 35, FI-40014, Jyv\"askyl\"a, Finland}%
\author{Jouni~Ruotsalainen}
\affiliation{Department of Physics, University of Jyv\"askyl\"a, P.O. Box 35, FI-40014, Jyv\"askyl\"a, Finland}%
\author{Marek~Stryjczyk}
\affiliation{Department of Physics, University of Jyv\"askyl\"a, P.O. Box 35, FI-40014, Jyv\"askyl\"a, Finland}%
\author{Ville~Virtanen}
\affiliation{Department of Physics, University of Jyv\"askyl\"a, P.O. Box 35, FI-40014, Jyv\"askyl\"a, Finland}%

\date{\today}
\begin{abstract}
A direct measurement of the ground-state-to-ground-state electron-capture decay $Q$ ($Q_{\rm EC}$) value of $^{97}$Tc has been conducted employing the high resolving power phase-imaging ion-cyclotron-resonance technique with the double Penning trap mass spectrometer JYFLTRAP. The resulting $Q_{\rm EC}$ value for $^{97}$Tc is 324.82(21) keV, exhibiting a precision approximately 19 times higher than the value adopted in the newest Atomic Mass Evaluation (AME2020) and differing by 1.2$\sigma$.
Furthermore, by combining this refined $Q$ value with nuclear energy-level data for the decay-daughter $^{97}$Mo, a potential ultra-low Q-value transition
$^{97}$Tc (9/2$^{+}$, ground state) $\rightarrow$ $^{97}$Mo$^{*}$ (320(1) keV), was detected.
The ground-state-to-excited-state electron-capture decay $Q$ value ($Q^{*}_{\rm EC}$) of this transition was determined to be 4.8(10) keV, confirming it to be energetically allowed with a confidence level of exceeding 4$\sigma$. 
The captures of electrons occupying the L and higher shells for this transition  are energetically allowed, giving a value of 2.0(10) keV for the closest distance of $Q^{*}_{\rm EC}$ to the allowed binding energy of the L1 shell.  
To predict partial half-lives and energy-release distributions for this transition, the atomic self-consistent many-electron Dirac--Hartree--Fock--Slater method and the nuclear shell model have been employed. 
Dominant correction terms such as exchange and overlap corrections, as well as shake-up and shake-off effects, were included in the final results. Moreover, in the case of a possible allowed transition, the normalized distribution of released energy in the electron-capture decay of $^{97}$Tc to the excited 320-keV state of $^{97}$Mo, is compared with that of $^{163}$Ho, which is being used for electron-neutrino-mass determination.
A pseudo-experiment technique was introduced to calculate error propagation in half-life and the 68\% confidence interval for normalized energy distributions.
\end{abstract}
\maketitle
\section{Introduction}

Neutrinos are electrically neutral fermions that interact only via the weak force and gravity, and they  have an incredibly tiny rest mass compared to other elementary particles. Neutrinos do not participate in electromagnetic or strong interactions, thus allowing them to pass through matter unimpeded.
Up to now, three neutrino flavors have been found: electron neutrino ($\nu_{e}$), muon neutrino ($\nu_{\mu}$), and tau neutrino ($\nu_{\tau}$). For each neutrino, there also exists a corresponding antineutrino, which also has spin of 1/2 and no electric charge~\cite{Petcov2013}.
Although neutrinos were initially thought to be massless as assumed by the standard model (SM), the three neutrino mass eigenstates do not show a one-to-one correspondence to the three flavors, each flavor being a quantum superposition of all three mass states~\cite{LESGOURGUES2006}.
Neutrinos oscillate between flavors, making their detection even more intriguing~\cite{Fukuda1998,SNOCollaboration2002,Gerbino2018a}. Neutrino-oscillation experiments reveal that at least two neutrino-mass eigenstates have a non-zero rest mass, leading to a mass difference between the three mass eigenstates. Laboratory experiments and cosmological observations have provided experimental data on differences in the squares of the neutrino masses, an upper limit on their sum, and an upper limit on the mass of the electron (anti)neutrino. Among these experiments, the most direct and model-independent way to pin down the exact mass value of the electron antineutrino (m$_{{\overline\nu_e}}$) is to measure the electron energy spectrum of single $\beta^{-}$ decay, e.g., the Karlsruhe Tritium Neutrino (KATRIN) experiment~\cite{Drexlin2013,Aker2019,Aker2022}, which has recently pushed the absolute electron antineutrino mass scale boundary to 0.45 eV/c$^{2}$ at the 90\% confidence level (CL)~\cite{KATRIN2025}.
KATRIN will finally set the  m$_{{\overline\nu_e}}$ upper limit  with a sensitivity of 0.2 eV/c$^{2}$ at 90\% CL. Other single-$\beta^{-}$-decay experiments, Project 8 and PTOLEMY project, using cyclotron radiation emission spectroscopy and the development of atomic tritium sources, will eventually allow a sensitivity of tritium-based neutrino-mass experiments to m$_{{\overline\nu_e}}$ beyond the design sensitivity of KATRIN~\cite{Esfahani2017,Betti2019} down to 0.04 eV/c$^{2}$. An alternative method, employing electron capture in $^{163}$Ho (ECHo), an experiment to calorimetrically measure the electron-capture (EC) deexcitation spectrum~\cite{Gastaldo2014,Gastaldo2017,Velte2019}, has set a current limit of 150 eV/c$^{2}$ for the electron neutrino mass~\cite{Velte2019}.

In single-$\beta^{-}$-decay experiments aimed at determining the mass of the electron antineutrino, a $Q$ value of the decay as small as possible is sought. A smaller $Q$ value leads to a higher fraction of effective decays within a given energy interval $\Delta{E}$ near the endpoint region. Currently, the running direct (anti)neutrino-mass-determination experiments rely exclusively on ground-state-to-ground-state (gs-to-gs) decay cases, such as $^{3}$H ($\beta^-$ decay) and $^{163}$Ho (EC). However, ongoing intensive searches are actively exploring isotopes having low $Q$-value ground-state-to-excited-state (gs-to-es) $\beta^-$/EC decays \cite{115Sn_2007,Haaranen2013,Suhonen2014,Ejiri2019} using high-precision Penning-trap mass spectrometry (PTMS) including JYFLTRAP, LEBIT, CPT, ISOTRAP, and SHIPTRAP Penning traps~\cite{Sandler2019,Karthein2019a,DeRoubin2020,ge2021,ge2021b,Ge2022a,ERONEN2022,Ge2022a,Ge2022b,Ramalho2022,Gamage22,Keblbeck2023,Ge2023,valverde2024,Ge2024,Ge2024b}. PTMS is the most precise and accurate technique for determining atomic masses and $Q$ values. It remains the only direct method capable of measuring decay $Q$ values to sub-keV precision or better, allowing verification of potential ultra-low (< 1 keV) $Q$-value transitions. 

For the nuclide $^{97}$Tc, there exists a potential low $Q$-value gs-to-es EC transition, $^{97}$Tc (9/2$^{+}$, ground state) $\rightarrow$ $^{97}$Mo$^{*}$ (320(1) keV~\cite{NNDC}, excited state), that could be employed for neutrino-mass detection. This speculated allowed transition, having a gs-to-es $Q$-value of 0.0(41) keV deduced from the existing experimental data~\cite{NNDC,Wang2021}, makes itself a candidate for an ultra-low $Q$-value transition. However, the $Q$ value of the decay to the excited state in the daughter nucleus is known with only 4.1 keV precision. This uncertainty arises from the combined uncertainties in the gs-to-gs EC $Q$ value (4 keV) adopted from~\cite{Wang2021}  and in the excitation energy (1 keV) from~\cite{NNDC}. As a result, a definitive conclusion regarding its energetic feasibility cannot be drawn solely from the current experimental information.
To further determine whether this specific decay is energetically feasible and whether it falls into the ultra-low category, experimental measurements of the gs-to-gs decay $Q$ value and the excitation energy of the state in the daughter nucleus are required.

Here, we present the first direct measurement of the gs-to-gs EC $Q$ value for $^{97}$Tc using the JYFLTRAP Penning-trap mass spectrometer. The precise $Q$ value obtained, along with current nuclear energy-level information for excited states of $^{97}$Tc, is employed to refine the gs-to-es $Q$ values. The possible ultra-low $Q$-value transition in $^{97}$Tc is verified with the newly determined experimental data. We confirm the transition to be energetically allowed. 
To explore the viability of this transition for neutrino-mass detection, we utilize two computational methods: the atomic self-consistent many-electron Dirac--Hartree--Fock--Slater method and the nuclear shell model. These approaches enable us to predict the partial half-lives and energy-release distributions for the relevant EC-decay transition.

\section{Experimental method}

\begin{figure}[!htb]
\centering
\includegraphics[width=0.95\columnwidth]{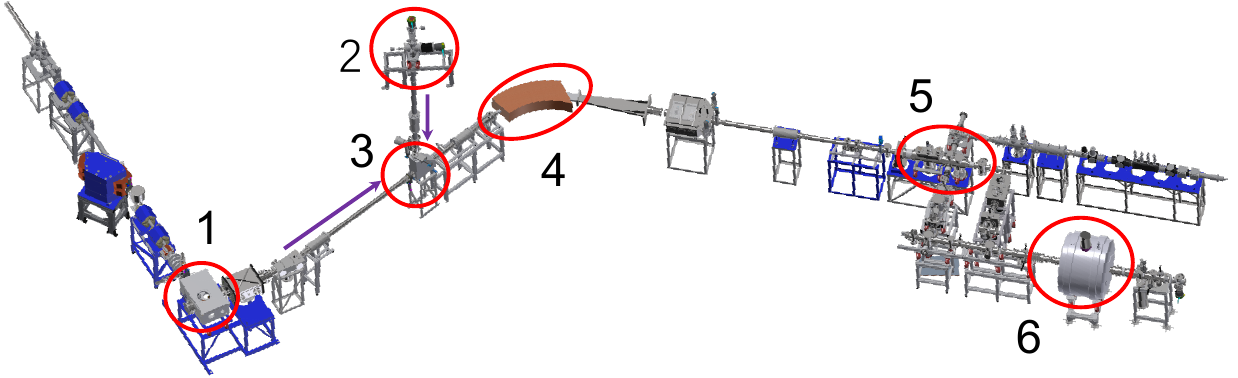}
\caption{(Color online).  Schematic view of the IGISOL facility. The $^{97}$Tc$^{+}$ ions were generated via proton-induced fusion-evaporation reactions on a natural Mo target in the IGISOL target chamber (1). Simultaneously, stable $^{97}$Mo$^{+}$ ions were created at the offline ion source station (2) using glow discharge. An electrostatic kicker (3) was utilized for ion-beam selection. A dipole magnet (4) selected singly-charged ions with the same mass number. The RFQ-CB (5) was employed for ion cooling and bunching, and the final $Q$-value and mass measurements were performed with the JYFLTRAP Penning-trap setup (6).}
\label{fig:igisol}
\end{figure}

The experiment was carried out at the Ion Guide Isotope Separator On-Line facility (IGISOL) at the University of Jyv\"askyl\"a, Finland~\cite{Moore2013,Kolhinen2013}.  
Figure~\ref{fig:igisol} gives a schematic view of the experimental setup.
 
$^{97}$Tc ions were generated in fusion-evaporation reactions between a natural molybdenum target foil irradiated with a few $\mu$A of a proton beam of energy 45 MeV produced by the K-130 cyclotron. The secondary ions  were stopped in a small volume gas cell filled with helium at a pressure of around 100 mbar as shown in Fig.~\ref{fig:igisol}. 
There, the ions end up mainly singly charged from charge-exchange reactions. The predominantly singly charged ions are extracted from the gas cell by helium gas flow and are guided with electric fields facilitated by differential pumping and a sextupole ion guide (SPIG)~\cite{Karvonen2008}.
Subsequently, the ions were accelerated using a 30 kV electric potential and transported to the 55$^\circ$ dipole magnet for isobaric mass separation with a typical mass resolving power of approximately 500. The isobarically separated ions with mass-to-charge ratio $A/q$ = 97 (including reaction products such as $^{97}$Nb$^{+}$, $^{97m}$Tc$^{+}$, $^{97}$Tc$^{+}$, and $^{97}$Mo$^{+}$) were then directed to a radiofrequency-quadrupole cooler-buncher (RFQ-CB)~\cite{Nieminen2001}, where they underwent accumulation, cooling, and bunching.
To produce the decay-daughter ion $^{97}$Mo$^{+}$ samples, an offline glow-discharge ion source was used \cite{Vilen2020a}.
Additionally, a 90$^\circ$ electrostatic bender, as presented in Fig.~\ref{fig:igisol}, selectively guided ions either from the online target station or the offline ion source for downstream transmission to the RFQ-CB.

The cooled and bunched ions from the RFQ-CB, were further directed to the JYFLTRAP~\cite{Eronen2012a}, which consists of two cylindrical Penning traps within a 7-T superconducting solenoid, for the final mass and ${Q}$-value measurements. The first trap, filled with helium buffer gas, serves as a purification trap and is employed for isobaric purification using the sideband buffer gas cooling technique~\cite{Savard1991} with a resolving power of approximately $10^{5}$.
To select the $^{97}$Tc$^{+}$ ion samples, this method could separate the aforementioned ions of $^{97}$Nb$^{+}$, but lacked sufficient mass resolving power to eliminate $^{97}$Mo$^{+}$ and $^{97m}$Tc$^{+}$ ions.

Mono-isotopic samples of $^{97}$Tc$^{+}$ were therefore prepared using the coupling of the dipolar excitation with Ramsey’s method of time-separated oscillatory fields~\cite{ERONEN20084527} and the phase-imaging ion-cyclotron-resonance (PI-ICR) technique~\cite{nesterenko2021,Nesterenko2018}, as described in~\cite{Ge2023}.
A Ramsey-type dipole excitation frequency scan in the second trap, employing a 2 ms (On) - 98 ms (Off) - 2 ms (On) excitation pattern, was used. The ions of interest were selectively chosen by filtering through positional gates using PI-ICR identification with a 436 ms phase accumulation time.

The PI-ICR method~\cite{Nesterenko2018,nesterenko2021} is used to measure the $Q$ value via the actual measurements of the cyclotron frequency, 
\begin{equation}
    \nu_c = \frac{1}{2\pi}\frac{q}{m}B,
\end{equation}
where $m$ is the mass of the ion, $q$ the charge and $B$ the magnetic field of the trap. 
The PI-ICR technique is around 25 times faster  than the conventional  time-of-flight ion-cyclotron-resonance (TOF-ICR) method used to achieve the same precision with a similar experimental condition~\cite{Nesterenko2018,nesterenko2021,Eliseev2014,Eliseev2013}. 
The scheme of the PI-ICR technique used at JYFLTRAP~\cite{Nesterenko2018} relies on the direct measurements of the magnetron motion and the reduced cyclotron motion at the same time while projecting the radial ion motion onto a position-sensitive MCP detector. 
The angle $\alpha_c$ between the two phase images of the projected radial motions, obtained from two timing patterns for determining the cyclotron frequency $\nu_c$, is defined with respect to the center spot: $\alpha_c = \alpha_+ - \alpha_-$. Specifically, $\alpha_+$ and $\alpha_-$ correspond to the polar angles of the magnetron and reduced cyclotron motion phases.
The cyclotron frequency $\nu_{c}$ is determined from the $\alpha_c$ measurements during the phase accumulation time $t_{acc}$: 
\begin{equation}
\label{eq:nuc2}
\nu_{c}=\frac{\alpha_{c}+2\pi n_{c}}{2\pi{t_{acc}}},
\end{equation}
where $n_{c}$ represents the full number of revolutions of the measured ions. 
To unambiguously assign $n_c$ in Eq.~\eqref{eq:nuc2}, different accumulation times were used for $^{97}$Tc$^{+}$. The actual measurements employed an accumulation time of 436 ms to determine the final cyclotron frequency $\nu_{c}$ for both $^{97}$Tc$^{+}$ and $^{97}$Mo$^{+}$ ions, ensuring that any leaked isobaric contaminant would not overlap with the ions of interest.

\begin{figure}[!htb]
   \includegraphics[width=0.9\columnwidth]{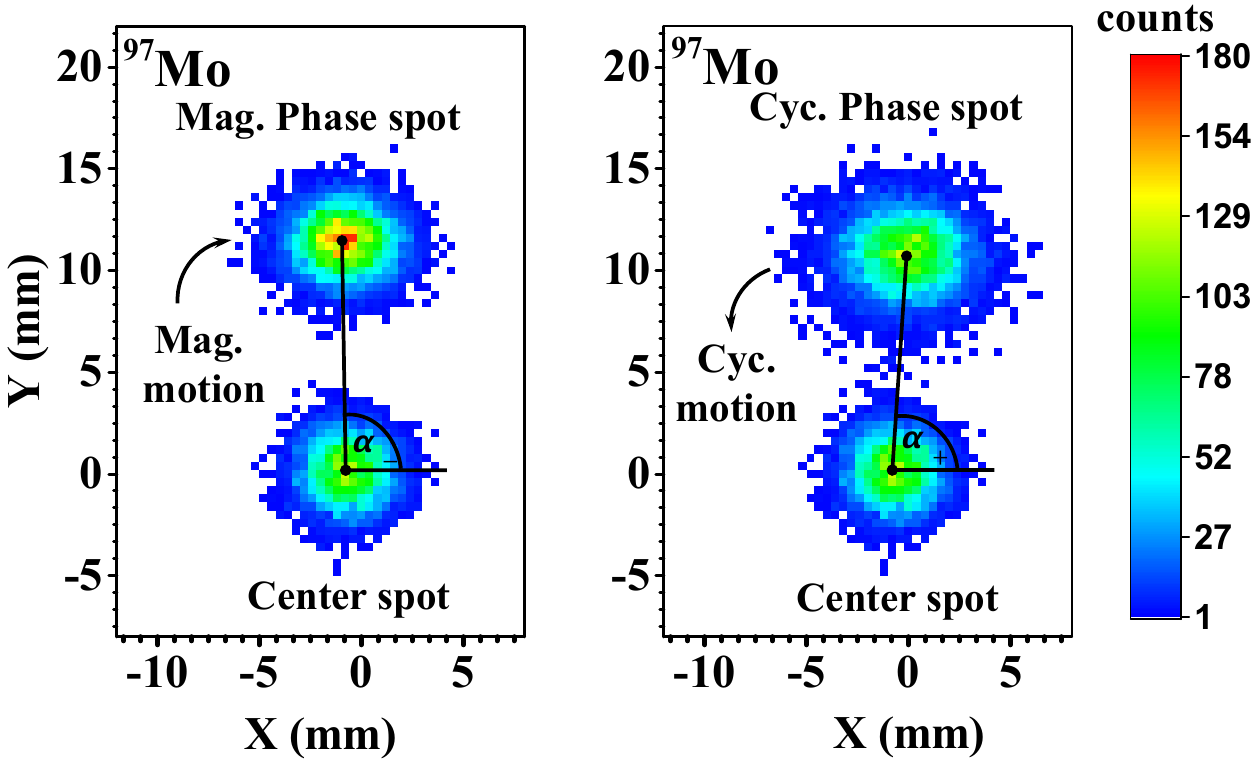}
   \caption{(Color online). Example $^{97}$Mo$^{+}$ ion spots of center, reduced cyclotron (Cyc.) phase and magnetron (Mag.) phase on the 2-dimensional position-sensitive MCP detector after a typical PI-ICR excitation pattern with an accumulation time of 436 ms. The Mag. phase spot is demonstrated on the left side and the Cyc. phase spot on the right. The angle difference between these two spots in relation to the center is employed to calculate the cyclotron frequency $\nu_{c}$. The color in each pixel corresponds to the number of ions as illustrated by color bars.
   }
   \label{fig:2-phases}
\end{figure}

To minimize any shifts in the $\nu_{c}$ ratio of the $^{97}$Tc$^{+}$-$^{97}$Mo$^{+}$ pair, the positions of the phase spots for the magnetron and reduced cyclotron motions were carefully chosen to maintain the angle $\alpha_c$ within a few degrees. This was done to account for the conversion of the reduced cyclotron motion to the magnetron motion and the potential distortion of the ion-motion projection onto the detector, ensuring that the shift remained well below 10$^{-10}$~\cite{Eliseev2014}. Figure~\ref{fig:2-phases} shows a representative measurement with "reduced cyclotron" and "magnetron" phase spots relative to the center spot. During the measurements, the excitation delay of $\nu_{+}$ was systematically scanned over one magnetron period, while the extraction delay varied over one cyclotron period, accounting for any residual motion that might have affected the spots. The interleaved measurements of $\nu_{c}$ for $^{97}$Tc$^{+}$-$^{97}$Mo$^{+}$ ions had a total data accumulation time of approximately 4.9 hours.

The gs-to-gs EC $Q$ value ($Q_{\rm EC}$), the mass difference of the decay pair, can be derived from their measured cyclotron frequency ratio $R$ (=$\frac{\nu_{c,d}}{\nu_{c,p}}$):
\begin{equation}
\label{eq:Qec}
Q_{EC}=(M_p - M_d)c^2 = (R-1)(M_d - qm_e)c^2+(R \cdot B_{d} - B_{p}),
\end{equation}
where $M_p$ and $M_d$ are the masses of the parent and daughter atoms, and $m_e$ the mass of an electron. The electron binding energies of the parent and daughter atoms, denoted as $B_p$ and $B_d$, are neglected due to their small values (on the order of a few eVs~\cite{NIST_ASD}), and $R$ is close to 1.
Given that both the parent and daughter ions share the same $A/q$ value and their relative mass difference $\Delta M/M < 10^{-4}$, the mass-dependent error becomes insignificant compared to the statistical uncertainty achieved in the measurements. Additionally, the contribution of uncertainty to the $Q$ value arising from the mass uncertainty of the daughter ion chosen as a reference, 0.16 keV/c$^2$ for $^{97}$Mo~\cite{Wang2021}, can be safely disregarded.

\begin{table}[!htb]
\caption{Final results from the analysis of the mean cyclotron frequency ratio between the ground state of daughter  $^{97}$Mo and parent ($^{97}$Tc,  9/2$^{+}$) nuclei.  Frequency ratio $\overline R$, $Q_{EC}$  values (in keV) and the mass excess (in keV/c$^2$) of parent nuclei determined in this work are listed in comparison with the AME2020 values~\cite{Wang2021}.}
\begin{ruledtabular}
   \begin{tabular*}{\textwidth}{cccc}
 & $\overline{R}$ & $Q_{EC}$ & \makecell[c]{mass excess \\}\\
\hline\noalign{\smallskip}
$^{97}$Tc (AME2020) &  & 320.0(40)&-87224.0(40) \\
$^{97}$Tc (This Work) &  1.00 000 359 84(23) &  324.82(21)& -87219.88(26) \\
   \end{tabular*}
   \label{table:Q-value}
\end{ruledtabular}
\end{table}

\begin{table*}[!htb]
   \caption{Potential candidate transitions from the initial state (ground state) of the parent  nucleus $^{97}$Tc (9/2$^{+}$) to the final states in the daughter nucleus $^{97}$Mo. The first column lists the excited final state of  $^{97}$Mo for the low $Q$ value transition. The decay type is provided in the second column. The third to sixth columns present the derived decay $Q_{\rm EC}$ values, sourced from literature (Lit.)~\cite{Wang2021} and this work (T. W.), respectively. 
   The seventh column displays the experimental excitation energy $E^{*}$~\cite{NNDC}.  The eight column shows the confidence level ($\sigma$) of the $Q^*_{\rm EC}$ being positive.  Columns nine to eleven, denoted as $\Delta{x}$, represent the distance of $Q_{\mathrm{EC}}^{}$ values to the binding energy $\varepsilon_x$ of the electrons in the daughter atoms, sourced from~\cite{X-Ray_Data_Booklet}. 
   Spin-parity assignments and energy values enclosed in braces \{\} signify uncertain assignments or uncertainties in excitation energy, resulting in uncertainties in the decay type or decay energy. All decay-energy and energy-level values are in units of keV.
   }
   \begin{ruledtabular}
   \begin{tabular*}{\textwidth}{ccccccccccc}
Final state 
& \makecell[c]{$Q_{\rm gs-gs}$ \\(Lit.)} & \makecell[c]{$Q_{\rm gs-gs}$ \\(T. W.)} &\makecell[c]{$Q^*_{\rm EC}$ \\(Lit.)} &\makecell[c]{$Q^*_{\rm EC}$ \\ (T. W.)}& $E^{*}$ & \makecell[c]{$Q/\delta Q$ \\(T. W.)} &\makecell[c]{$\Delta_{\mathrm{L1}}$ \\ (T. W.)}  &\makecell[c]{$\Delta_{\mathrm{L2}}$ \\ (T. W.)}  & \makecell[c]{$\Delta_{\mathrm{M1}}$ \\ (T. W.)} \\
\hline\noalign{\smallskip}
   $^{97}$Mo ($1/2:9/2^{+}$) &320.0(40)&324.82(21)& 0.0(41)  & 4.8(10) & 320.0(10)& 4.8 & 2.0(10) &2.2(10)& 4.3(10)  \\
   $^{97}$Mo (5/2$^{+}$, ground state)  &320.0(40)&324.82(21)&   & && & &&   \\
   \end{tabular*}
   \label{table:low-Q}
   \end{ruledtabular}
\end{table*}

\section{Results and discussion}
\begin{figure}[!htb]
   \includegraphics[width=0.99\columnwidth]{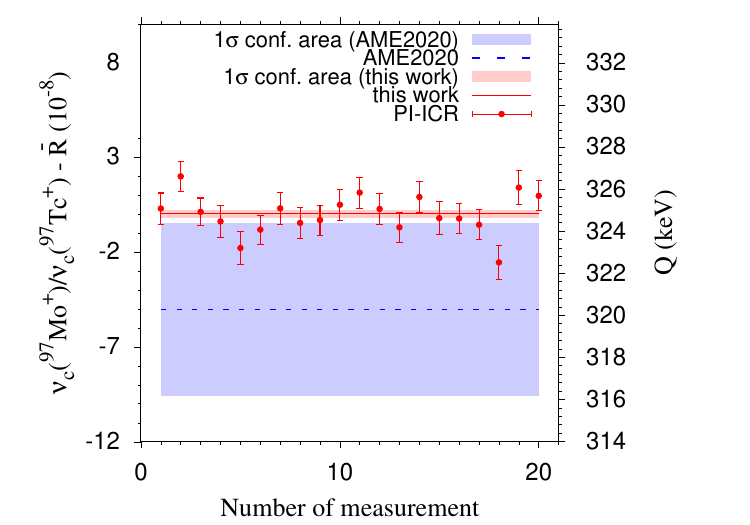}
   \caption{(Color online). The measured experimental results from this work compared to the literature values~\cite{Huang2021,Wang2021}. The deviations of the individually measured cyclotron frequency ratios  $R$ ($\nu_c$($^{97}$Mo$^{+}$)/$\nu_c$($^{97}$Tc$^{+}$)) from the measured value $\overline{R}$ (left axis) and $Q$ value (right axis) in this work are compared to values adopted from AME2020. The red points with uncertainties represent individual measurements using the PI-ICR method. Vertical brown dashed lines separate measurements conducted at different time slots. The weighted average value $\overline{R}$ (referenced in Table~\ref{table:Q-value}) is depicted by the solid red line, and its 1$\sigma$ uncertainty band is shaded in red. The dashed blue line illustrates the difference between our new value and the one referenced in AME2020, with its 1$\sigma$ uncertainty area shaded in blue.}
   \label{fig:ratio}
\end{figure}

From the measurements of the cyclotron  frequency ratio $R$,  $Q_{\rm EC}$ was calculated via Eq.~\ref{eq:Qec}.  
As shown in Fig.~\ref{fig:ratio}, 20 PI-ICR measurement data sets were collected for $^{97}$Tc$^{+}$-$^{97}$Mo$^{+}$. 
In less than 5 minutes, a full scanning measurement (one cycle) of the magnetron phase, reduced cyclotron phase and center spot in sequence  was completed for each ion species of the decay pair $^{97}$Tc$^{+}$-$^{97}$Mo$^{+}$. 
In the analysis, we determined the position of each spot using the maximum likelihood method. 
After summing a few cycles to obtain reasonable counts for fitting, phase angles were calculated based on the determined positions of the phase spots as demonstrated in Fig.~\ref{fig:2-phases}. These phase angles allowed us to deduce the cyclotron frequencies of each ion species. 
By linearly interpolating the $\nu_{c}$ of the daughter ion $^{97}$Mo$^{+}$ (reference) to the time of the measurement of the parent ion $^{97}$Tc$^{+}$ (ion of interest),  the cyclotron frequency ratio $R$ was derived. 
To minimize potential cyclotron frequency shifts resulting from ion-ion interactions~\cite{Kellerbauer2003,Roux2013}, a median ion rate was constrained to 1-2 ions per bunch during the measurements. Bunches with fewer than five detected ions per bunch were considered during the data analysis. In the analysis, the  count-rate related frequency shifts were not observed.
The temporal fluctuation of the magnetic field  $\delta_B(\nu_{c})/\nu_{c}=  \Delta t \times 2.01 \times  10^{-12}$/min~\cite{nesterenko2021} is taken in account in the final results.
Here, $\Delta t$ represents the time interval between consecutive reference measurements. 
The contribution of magnetic field fluctuations to the final frequency ratio uncertainty was less than 10$^{-10}$, as the parent-daughter measurements were interleaved with intervals less than 10 minutes. Additionally, the frequency shifts in the PI-ICR measurement caused by ion image distortions were ignored, as they were well below the statistical uncertainty. Furthermore, the fact that decay pair ions $^{97}$Tc$^{+}$-$^{97}$Mo$^{+}$ are mass doublets, cancels out many systematic uncertainties in the cyclotron frequency ratio. 
The weighted mean ratio $\overline{R}$ of all single frequency ratios was calculated, considering the inner and outer errors to deduce the Birge ratio~\cite{Birge1932}. The maximum of the inner and outer errors served as the weight for computing $\overline{R}$. Results of our analysis including all data sets, in comparison to literature values, are presented in Fig.~\ref{fig:ratio} and Table~\ref{table:Q-value}. 
The final daughter-to-parent frequency ratio $\overline{R}$ with its uncertainty is determined to be 1.000 003 598 4(23), resulting in a gs-to-gs $Q$ value of 324.82(21) keV. 

\begin{figure} [!htb] 
   \includegraphics[width=0.78\columnwidth]{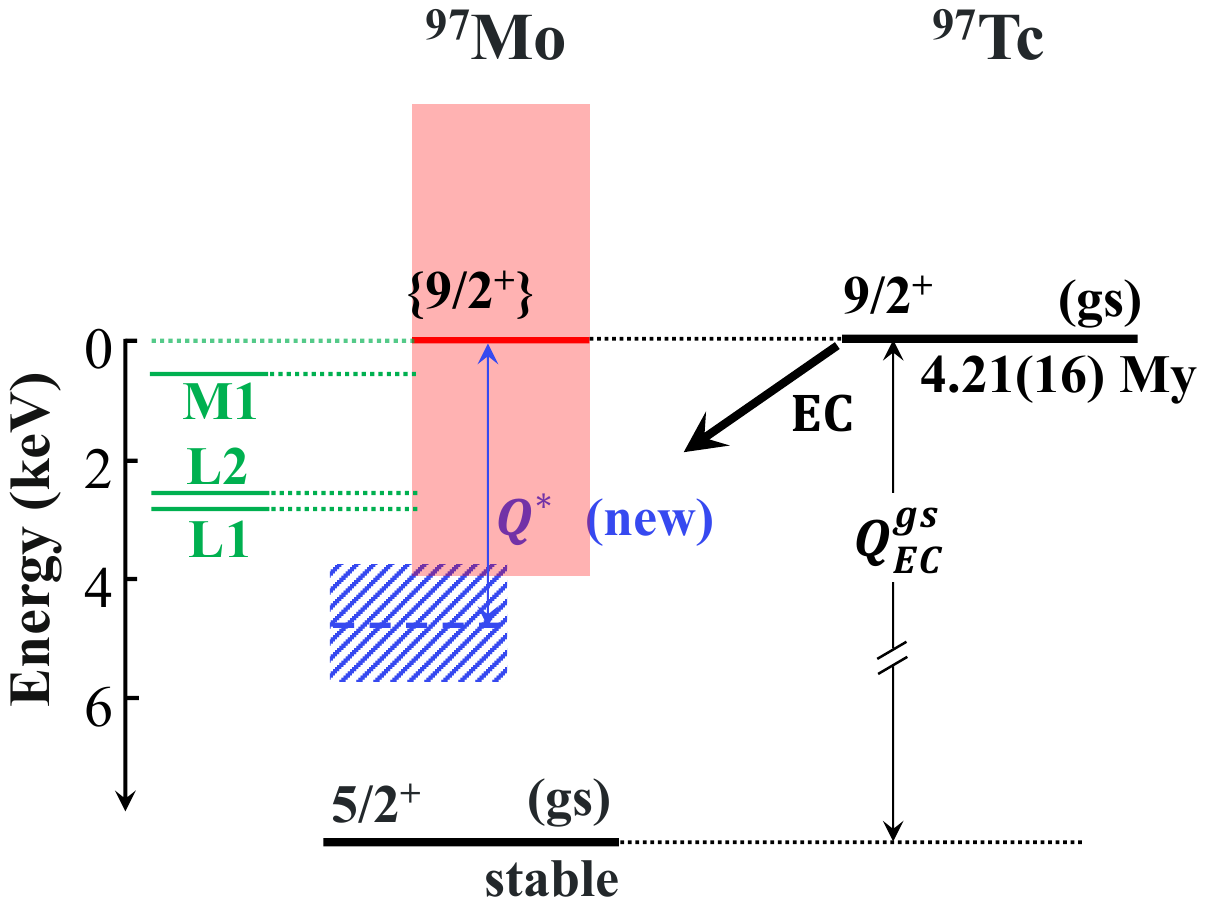}
   \caption{(Color online). Partial EC 
   scheme of $^{97}$Tc decaying to $5/2^+$ ground and $\{9/2^+\}$ final states, the spin-parity assignment being uncertain for the latter state. The dashed green horizontal lines extending from the left indicate the threshold energy for the marked electron orbitals from which the EC can happen. The boxes with a red horizontal line show the $Q_{\mathrm{EC}}$ value using data from literature~\cite{Huang2021,Wang2021} with 1$\sigma$ uncertainty in comparison to the $Q$ value from this work (hatched blue box). 
   Our results indicate that EC to the $\{9/2^+\}$ state can occur only from the L1 orbital or higher, while the data from literature cannot give a conclusive clue of whether there are any allowed electron orbitals higher than the K orbital for EC to occur. 
   }
   \label{fig:level-sheme}
\end{figure}

The gs-to-gs $Q_{\rm EC}$ value of 324.82(21) keV obtained from this work, which represents the first direct measurement, is approximately 19 times more precise than the value deduced from the evaluated masses in AME2020
~\cite{Huang2021,Wang2021}. The measured  $Q_{\rm EC}$ value has a deviation of 4.8(40) keV from the AME2020 value and is $\approx$1$\sigma$ larger.

In the case of EC, the closer is the $Q$ value of the decay to one of the ionization energies of the captured electrons, the larger the resonance enhancement of the rate near the end-point, where the effects of a non-vanishing neutrino mass are relevant. 
Which orbital electrons take part in the EC process and the absolute $Q$ values of the decays are crucial for modelling the spectrum shape near the endpoint.
The current precision deduced from literature~\cite{NNDC,Huang2021,Wang2021} does not allow a definitive conclusion on whether any other electron orbitals allow EC to occur for the transition $^{97}$Tc (9/2$^+$) $\rightarrow$ $^{97}$Mo$^*$ (320.0 keV).
In this work, the high-precision $Q$-value measurement allows an unambiguous characterization of all the possible lines in the EC spectrum at a significance level of at least 2$\sigma$ for the transition. This makes the modeling of its shape possible.
The high-precision EC energy measured from this work, together with the nuclear energy level data from Ref.~\cite{NNDC} of the excited state of $^{97}$Mo as tabulated in Table~\ref{table:low-Q},  was used to determine the gs-to-es $Q$ value ($Q^*_{\rm EC}$) of the candidate state.  The newly determined $Q^*_{\rm EC}$ confirms that the decay of the ground state of $^{97}$Tc to the  excited state of interest is energetically allowed with a confidence level of more than 4$\sigma$, as shown in Table~\ref{table:low-Q}. 
A comparison of the $Q$ values of the gs-to-es EC transition from this work to the value derived from AME2020 is shown in Fig.~\ref{fig:level-sheme}. 
As tabulated in Table~\ref{table:low-Q}, $\Delta_{x}$  gives the distance of the $Q_{\mathrm{EC}}^{*}$  value to the binding energy $\varepsilon_x$~\cite{X-Ray_Data_Booklet} of the electrons in the allowed daughter atomic shells ($x$ = L1, L2, and other electrons from s levels and p$_{1/2}$ levels from the third and higher shells). 
For the state with the excitation energy of 320.0(10) keV, the captures of electrons occupying the L1 and higher shells for the transition $^{97}$Tc (9/2$^+$) $\rightarrow$  $^{97}$Mo$^*$ are energetically allowed as indicated in Fig.~\ref{fig:level-sheme}.
The transition  $^{97}$Tc (9/2$^+$) $\rightarrow$ $^{97}$Mo$^*$ (320.0 keV), giving the values of 2.0(10) keV and 2.2(10) keV for the distance of $Q_{\mathrm{EC}}^{*}$ to the allowed binding energies $\varepsilon_{L1}$ =  2.866 keV and $\varepsilon_{L2}$ = 2.625 keV \cite{X-Ray_Data_Booklet}, is potentially of the allowed Fermi plus Gamow-Teller type. This transition is of high interest for future neutrino mass determination due to a possible high branching ratio.  To confirm whether the emitted neutrino energy 2.0(10) keV is ultra-low, further high-precision measurements of the excitation energy of the state are required. Moreover, the parity of this state needs to be determined to verify the decay type of the transition to this state.

\section{Theoretical predictions}

Two calculation methods are employed to predict the distribution of energy released in the decay, the transition half-life and their error spread, namely the atomic many-electron Dirac--Hartree--Fock--Slater (DHFS) self-consistent method and the Nuclear Shell Model (NSM) many-nucleon framework. 
For the atomic-structure calculations we made use of the RADIAL subroutine package \cite{SalvatCPC2019}, which also contains the \textsc{DHFS.F} code, while for the NSM we employed the NuShellX\@MSU code \cite{Brown2014}.

In our calculation we consider the initial atom in its ground state. The electronic structure of the final atom is constructed starting from the configuration of the initial atom with an electron hole in the shell from where it was captured.
When the nucleus captures the electron, the other electrons (the spectator electrons) undergo processes that impact the decay rate. A couple of important corrections related to these processes are considered \cite{Ge2024b}:  
exchange and overlap corrections, shake-up and shake-off effects.

The atomic shells are denoted using the main quantum number $n$, and the relativistic quantum number $\kappa$ as $x=(n,\kappa)$. In order to compute the released energy distribution of an allowed EC event we sum all the contributions from the atomic shells with the relativistic quantum number $\kappa=\pm1$ as:
\begin{equation}
\label{eq:rho}
\rho(E^{\prime})=\frac{G_{\beta}^{2}}{(2 \pi)^{2}} \sum_{x} n_{x} C_{x} \beta_{x}^{2} B_{x} S_{x} p_{\nu} E_{\nu} \frac{\Gamma_{x} /(2 \pi)}{\left(E^{\prime}-\varepsilon_{x}\right)^{2}+\Gamma_{x}^{2} / 4}.
\end{equation}

\begin{table*}[!htb]
   \caption{Computed half-lives for the EC decay of $^{97}$Tc to the excited state (320.0 keV) in $^{97}$Mo, with the $Q^*_{\rm EC}$ value shown in Table \ref{table:low-Q}. The errors are computed according to the approach presented in Sec. \ref{statistical}. The used interaction is jj45pnb. LCS denotes Lowest Captured Shell}
  \begin{ruledtabular}
  \begin{tabular*}{\textwidth}{lc|ccccccccc}
 LCS & Decay & Total half-life &$\mathrm{~L} 1$ & $\mathrm{~L} 2$ & $\mathrm{M} 1$ & $\mathrm{M} 2$ & $\mathrm{N} 1$ & $\mathrm{N} 2$ & $\mathrm{O} 1$ &\\
 & type & $(10^4 \textrm{yr})$ & $(10^4 \textrm{yr})$ & $(10^7 \textrm{yr})$ & $(10^4 \textrm{yr})$ & $(10^6 \textrm{yr})$ & $(10^5 \textrm{yr})$ & $(10^7 \textrm{yr})$ & $(10^6 \textrm{yr})$ &\\
\hline\noalign{\smallskip}
L1 & allowed & $19^{+17}_{-0.7}$ & $ 46^{+120}_{-26}$ & $2.2^{+5.0}_{-1.2}$ & $ 47^{+32}_{-17}$ & $26^{+18}_{-10}$ & $20^{+12}_{-7}$ & $ 13^{+8}_{-4}$ & $42^{+25}_{-15}$ \\


L2 & allowed & $93^{+7}_{-6}$ & $-$ & $570^{+4000}_{-340}$ & $ 169^{+50}_{-34}$ & $91^{+27}_{-18}$ & $60^{+18}_{-12}$ & $ 38^{+11}_{-8}$ & $127^{+40}_{-25}$ \\


\hline

& & Total half-life &$\mathrm{L} 1$ & $\mathrm{L} 2$ & $\mathrm{L} 3$ & $\mathrm{M} 1$ & $\mathrm{M} 2$ & $\mathrm{M} 3$ & $\mathrm{M} 4$ & $\mathrm{M} 5$ \\
& & $(10^{16}\textrm{yr})$ & $(10^{23}\textrm{yr})$ & $(10^{24}\textrm{yr})$ & $(10^{18}\textrm{yr})$ & $(10^{21}\textrm{yr})$ & $(10^{23}\textrm{yr})$ & $(10^{18}\textrm{yr})$ & $(10^{21}\textrm{yr})$ & $(10^{16}\textrm{yr})$ \\

\hline

L1 & 2nd UF & $11^{+6.1}_{-3.4}$ & $1.0^{+50}_{-0.92}$ & $3.1^{+84}_{-2.7}$ & $5.3^{+38}_{-4.0}$ &
$4.4^{+15}_{-3.1}$ & $2.3^{+7.2}_{-1.6}$ & $2.0^{+3.1}_{-1.1}$ & $1.1^{+1.7}_{-0.62}$ & $11^{+6.2}_{-3.4}$
\\

L2 & 2nd UF & $33^{+2.5}_{-1.9}$ & $-$ & $4.9^{+2326}_{-4.5}\times10^{7}$ & $4.1^{+21}_{-2.7}\times10^{4}$ &
$215^{+59}_{-38}$ & $98^{+26}_{-17}$ & $24^{+4.0}_{-2.8}$ & $12^{+1.9}_{-1.4}$ & $35^{+2.6}_{-2.0}$
\\

\hline
& & Total half-life &$\mathrm{L} 1$ & $\mathrm{L} 2$ & $\mathrm{L} 3$ & $\mathrm{M} 1$ & $\mathrm{M} 2$ & $\mathrm{M} 3$ & $\mathrm{M} 4$ & $\mathrm{M} 5$ \\
& & $(10^{21}\textrm{yr})$ & $(10^{28}\textrm{yr})$ & $(10^{30}\textrm{yr})$ & $(10^{24}\textrm{yr})$ & $(10^{26}\textrm{yr})$ & $(10^{28}\textrm{yr})$ & $(10^{23}\textrm{yr})$ & $(10^{26}\textrm{yr})$ & $(10^{21}\textrm{yr})$ \\

\hline

L1 & 4th NUF & $10^{+10}_{-3.6}$ &$5.7^{+1066}_{-5.5}$& $1.4^{+116}_{-1.3}$&$2.6^{+58}_{-2.3}$ &
$5.0^{+31}_{-4.0}$ & $2.5^{+1.4}_{-2.0}$ & $2.6^{+8.4}_{-1.8}$ & $1.4^{+4.2}_{-0.10}$ & $7.2^{+11}_{-3.9}$
\\

L2 & 4th NUF & $74^{+11}_{-8.3}$& $-$ & $6.1^{+25320}_{-5.9}\times10^{9}$ & $1.9^{+26}_{-1.5}\times10^{6}$&
$900^{+340}_{-210}$ & $380^{+140}_{-87}$ & $110^{+29}_{-19}$ & $51^{+12}_{-8.4}$ & $78^{+12}_{-8.8}$

\end{tabular*}
   \label{table:T12-allowed}
   \end{ruledtabular}
\end{table*}

Here we employ two corrections, namely the exchange and overlap correction represented by the $B_{x}$ factor and the shake-up and shake-off corrections introduced as the $S_{x}$ factor \cite{MougeotARI2018}. The Coulomb amplitude is calculated for the initial atom and is represented by $\beta_{x}$. The Cabibbo angle $\theta_{\rm C}$ and the Fermi constant $G_{\rm F}$ are present in $G_\beta=G_{\rm F}\cos \theta_{\rm C}$. The relative occupancy of the shell is denoted as $n_{x}$. The energy and momentum of the emitted neutrino are represented as $E_{\nu}$ and $p_{\nu}=\sqrt{E_{\nu}^2-m_{\nu}^2}$, with $m_{\nu}$ being the neutrino mass. A recent improvement \cite{SevestreanPRA2023} in the computation of the relaxation energy following the capture is the usage of total binding energy as $\varepsilon_{x}=|T_{\rm g.s.}|-|T_{x}|$, where $T_{\rm g.s.}$  indicates the energy of the final atom in its ground state, while $x$ represents the energy of the excited state of the final atom with a hole in the $x$ shell. The $\Gamma_{x}$ represents the intrinsic line-widths of Breit--Wigner resonances centered at $\varepsilon_{x}$ and they were taken from \cite{CampbellADNDT2001}. 

The spin of the 320-keV excited state in $^{97}$Mo is uncertain, ranging within a range of $J_f=1/2^+ - 9/2^+$. This means that the decay transition from the $J_i=9/2^+$ ground state of $^{97}$Tc to the 320-keV state could be allowed ($J_f=9/2^+$ and $J_f=7/2^+$), second-forbidden non-unique (2nd NUF) in the case of $J_f=5/2^+$, second-forbidden unique (2nd UF) in the case of $J_f=3/2^+$ and fourth-forbidden non-unique (4th NUF) in the case of $J_f=1/2^+$. In the case of an allowed transition, the nuclear structure part is represented by the $C$ factor

\begin{equation}
C_{x}=\frac{1}{2J_i+1}\left(g_{\rm V}^2 \vert M_{\rm F}\vert^2 + g_{\rm A}^2 \vert M_{\mathrm{GT}}\vert^2\right),
\label{eq:Cx-allowed}
\end{equation}
with $g_{\rm V}$ being the weak vector coupling, and $g_{\rm A}$ and $J_i$ being the weak axial coupling and the angular momentum of the initial nucleus. The Fermi nuclear matrix element is denoted by $M_{\rm F}$ and the Gamow-Teller nuclear matrix element is denoted as $M_{\mathrm{GT}}$.

To compute the decay probability $\lambda$ we employ the formalism described in \cite{BambynekRMP1977}:

\begin{equation}
\lambda=\frac{G_{\beta}^{2}}{2 \pi^{3}} \sum_{x} n_{x} C_{x} f_{x} S_{x},
\end{equation}
where $f_{x}=\frac{\pi}{2} p_{\nu}^{2} \beta_{x}^{2} B_{x}$. For allowed transition we used the simplified approach for the $C_x$ factor in Eq. \ref{eq:Cx-allowed}, while for the unique and non-unique forbidden transitions, $C_x$ is defined as:

\begin{multline}
C_{x}=  {\left[M_{L}\left(k_{x}, k_{\nu}^{(1)}\right)+sign(\kappa_{x}) m_{L}\left(k_{x}, k_{\nu}^{(1)}\right)\right]^{2} } \\
 +\left[M_{L}\left(k_{x}, k_{\nu}^{(2)}\right)+sign(\kappa_{x}) m_{L}\left(k_{x}, k_{\nu}^{(2)}\right)\right]^{2} \\
 +\left[M_{L+1}\left(k_{x}, k_{\nu}^{(2)}\right)+sign(\kappa_{x}) m_{L+1}\left(k_{x}, k_{\nu}^{(2)}\right)\right]^{2}.
\end{multline}

The $M$ and $m$ factors are defined in \cite{BambynekRMP1977}.

\begin{figure}[!htb]
   \includegraphics[width=0.95\columnwidth]{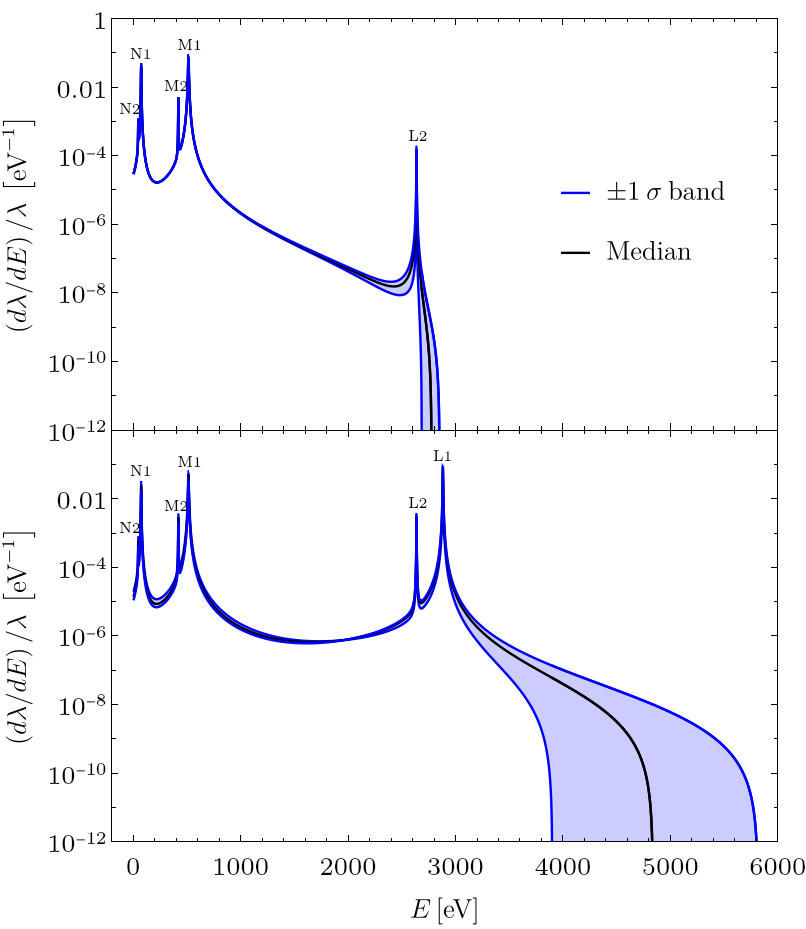}
   \caption{Normalized distributions of released energy as a function of $E$ in the statistical approach, Sec. \ref{statistical}. We represented the 68\% confidence interval shaded in blue with a dark blue contour. The blue lines indicate the 1$\sigma$ distributions. The black line represents the median value of the normalized energy distribution at each energy point. The plots are for EC decay of $^{97}$Tc to the excited state of interest in $^{97}$Mo with a $Q^{*}_{\rm EC}$= 4.8(10) keV assuming allowed transition type. The $Q$ value follows a broad Gaussian distribution. For 97.3$\%$ of the elements in the distribution, the L1 channel is energetically possible (bottom plot). For 1.2$\%$ of them, the L1 channel is energetically forbidden, but the L2 channel is possible (top plot).
   }
   \label{fig:5-68spectra}
\end{figure}

\begin{figure}[!htb]
   \includegraphics[width=0.98\columnwidth]{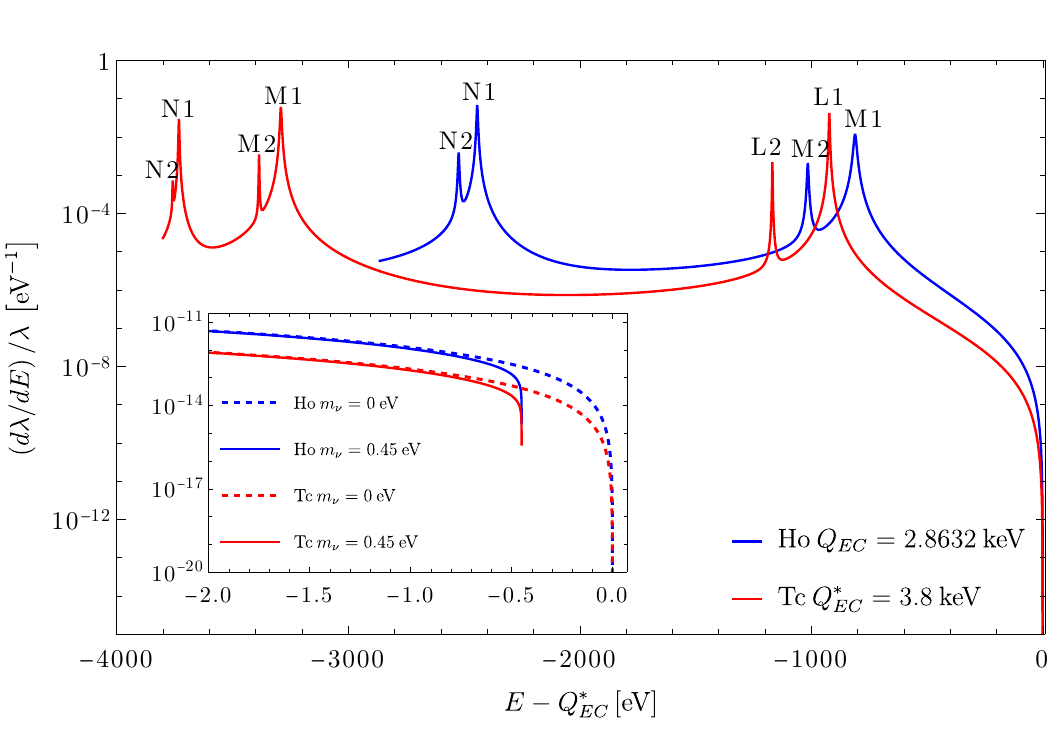}
   \caption{Normalized distributions of released energy in the EC decay of $^{97}$Tc in the transitions to the excited state of $^{97}$Mo, of interest here, as functions of $E-Q^{*}_{\rm EC}$, assuming allowed transition type. The blue line corresponds to the decay to the excited final state of energy $E^{*}=320.0$ keV and with  $Q^{*}_{\rm EC}=3.8$ keV, which is $1 \sigma$ away from the central value of $Q^{*}_{EC}=4.8$ keV. 
   L1, L2, M1, and N1 indicate subshells from which the electron was captured. The M2, N2 and O1 subshells are harder to distinguish and are not labeled. 
   }
   \label{fig:6-spectra}
\end{figure}

In Table \ref{table:T12-allowed}, we present the predicted half-lives and associated uncertainties for the decay of $^{97}$Tc to the excited state of $^{97}$Mo, corresponding to the experimental energy $E^{*}=320.0$ keV, across all relevant atomic shells. 
Due to the uncertainty of the angular momentum of the final state, we considered multiple possible values. By computing the energy levels using the jj45pnb interaction, we obtained a few possible values for J in the vicinity of 320 keV, namely 7/2, 3/2, and 1/2. Thus, we have the possible decay types allowed, 2nd unique forbidden, and 4th non-unique forbidden. The matrix elements needed for the half-life were computed with the same jj45pnb interaction. We also used the glekpn interaction for the allowed transition; the computed half-life was an order of magnitude smaller. Comparing the half-life for the excited state transition with the total half-life measured, $4.21\times10^{6}$ years \cite{NNDC}, one observes that for the allowed transition, the computed half-life is in tension with the measured one. As such, this is most likely due to the assumed angular momentum $J$ of the excited final state, which thus should be verified by new spectroscopic measurements. 
Considering the uncertainty in the $Q$ value, the capture from the L1 shell is uncertain but we retain this possibility in our computations.
However, the capture from the L2 shell is certainly possible. As can be seen, there is a significant difference between the upper and lower error because the $Q$-value spans a considerable interval of values, and it appears at various powers in the half-life formula.

In Fig. \ref{fig:5-68spectra} we present the normalized distributions of released energy within a shaded-in-blue 68$\%$ confidence band between the blue curves assuming an allowed transition. Considering a Gaussian distribution for the Q-value, 97.3$\%$ of cases permit the capture of an electron from the L1 shell (bottom plot). In 1.2$\%$ the L1 channel is energetically impossible but the capture from the L2 sub-shell is possible (top plot). The other 1.5$\%$ of cases were not presented here, due to their decreasing probability of occurrence. The black line represents the median values for each energy point in the graph. The interval was obtained as described in section \ref{statistical}. This analysis shows the high dependence of the normalized distribution on the value of $Q_{\mathrm{EC}}^{*}$, especially in the interesting region at the end of the spectrum. The greatest impact on the error of $Q_{\mathrm{EC}}^{*}$ comes from the uncertainty in the energy of the level of interest in the final nucleus. Thus a better assessment of the energy of this level is necessary in order to have a more well-defined distribution of energy at $Q^{*}_{\rm EC}$.

The current measurement of the value of $Q_{\mathrm{EC}}^{*}$ with the associated error does not give a definite picture of the capture process. 
In order to investigate the implications of the associated uncertainties in the case of an allowed transition, we present in Fig. \ref{fig:6-spectra} a comparison between the normalized distribution of the released energy as a function of the energy, $E-Q_{\mathrm{EC}}^{*}$, for the decay of 
$^{97}$Tc to the excited state of $^{97}$Mo with $Q_{\mathrm{EC}}^{*} = 3.8$ keV, within 1 $\sigma$ range of the experimental central value, and $^{163}$Ho to the ground state of $^{163}$Dy with $Q_{\mathrm{EC}} = 2.8623$ keV. As can be seen in the enlarged scale in the inset of Fig. \ref{fig:6-spectra}, the neutrino-mass-sensitive region provides a higher EC event rate for $^{97}$Tc for this Q value, making it comparable with the rate for $^{163}$Ho. Also, this inset shows exactly the impact of the neutrino mass on the shape of the spectrum and how it could be highlighted. 
A strong $^{97}$Tc source can be produced via the $(p,n)$ reaction with {19} MeV protons on an enriched $^{97}$Mo target (cross-section $\approx${206} mbarn from theoretical calculation with TALYS \cite{talys}). With a {10} mA beam and 1 month irradiation, we estimate producing $\sim 10^{19}$ atoms, sufficient for a neutrino mass experiment requiring sub-{keV} sensitivity over one year if the allowed transition can be confirmed. The long half-life of about 4 My for $^{97}$Tc renders it a good candidate for future long-term neutrino-mass measurements if the considered transition is of the allowed type. In case the transition is of forbidden type, it is not relevant for neutrino-mass measurements as witnessed by the very long half-lives of the transition in Table~\ref{table:T12-allowed}.

\subsection{Statistical approach \label{statistical}}
In order to calculate the propagation of errors in the half-life and the 68\% band in the normalized distributions of released energy as shown in Fig. \ref{fig:5-68spectra}, we employed the pseudo-experiment technique. This method involves sampling of parameters that potentially fluctuate from probability distribution functions (PDFs). For each set of parameters, the half-life times and the normalized distribution of released energy are computed, thus obtaining their PDFs. The central values are obtained as medians of the PDFs. For the two-sided bounds, we took the element at $0.16$ and $0.84$ of the CPDFs. The uncertainties are reported as the difference between each bound and the median value.

We took into account $Q^*_{\rm EC}$ and $g_{\rm A}$ as sources of uncertainty. Assuming a Gaussian distribution of $Q^*_{\rm EC}$, in about $2.7\%$ cases the capture from the L1 sub-shell will not be energetically possible. Thus, we split the sampled values into groups and considered two in the current study. For the first group, the capture of the L1 electron is possible (L1 allowed), while for the second, the L1 channel is energetically forbidden, and the L2 channel is energetically possible (L1 forbidden). For each group, we plotted the median and the 1$\sigma$ band of the normalized distribution of released energy.
For the distribution of $g_{\rm A}$ we have used a uniform distribution with values between $0.7$ and $1.0$. The error propagation based on $g_{\rm A}$ was taken into account only for the allowed transition.  We consider a number of $10^{6}$ pseudo-experiments. In particular, for the normalized distributions of released energy, we constructed a uniform grid of $10^{4}$ points for $E$. In each point, the procedure described above was applied.

\section{Conclusion and outlook}
A direct high-precision gs-to-gs  EC-decay $Q$-value measurement of $^{97}$Tc (9/2$^{+}$)$\rightarrow $$^{97}$Mo (5/2$^{+}$) was performed using the PI-ICR technique at the JYFLTRAP Penning-trap mass spectrometer.  A $Q$ value of 324.82(21)~keV  was obtained and the  precision  was improved by a factor of around 19 compared to literature. The measurement also improved the mass excess of $^{97}$Tc , -87219.88(26) keV/c$^2$, by a factor of 15 compared to the previously measured value.
With the refined gs-to-gs $Q$ value, the candidate transition of $^{97}$Tc (9/2$^{+}$)$\rightarrow $$^{97}$Mo$^{*}$ (320.0 keV) was validated to be energetically allowed. This also allows an unambiguous characterization of all the possible lines in the EC spectrum of this transition at a significance level of at least 2$\sigma$. 
The transition to the 320.0-keV state 
has a small distance of 2.0(1.0) keV for the gs-to-es $Q_{\mathrm{EC}}^{*}$  to the binding energy  of the electrons in the allowed daughter atomic shell L1. 
For further confirmation of whether this lowest emitted neutrino energy is ultra-low, the excitation energy needs to be determined with higher precision and accuracy. Also, to define the nature of the transition, allowed or forbidden, a spectroscopic measurement of the angular momentum of the final state should be performed.
A highly charged ion (HCI) Penning trap will refine the gs-to-gs $Q_{\mathrm{EC}}$ to a few eV \cite{Schweiger2024}, and cryogenic detectors, e.g., Transition Edge Sensors (TES) or Metallic Magnetic Calorimeters (MMCs) at low temperatures, will measure the 320.0~keV excitation energy to around 10s~eV, reducing $Q_{\mathrm{EC}}^{*}$ uncertainty to $\approx$10s~eV. Phase-space factor analysis and gamma coincidence further constrain rates, ensuring viability for sub-eV sensitivity, as with  $^{159}$Dy \cite{Gastaldo2022}. 
The atomic self-consistent many-electron Dirac--Hartree--Fock--Slater method and the nuclear shell model were utilized to predict the partial decay half-lives and energy distributions of gs-to-es EC transitions in $^{97}$Tc with low $Q$ values. To calculate error propagation in half-life and the 68\% confidence interval for normalized energy distributions, a pseudo-experiment technique was introduced. Multiple corrections, such as exchange, overlap, shake-up, and shake-off effects, were accounted for in these predictions for modeling the spectrum shape near the endpoint.


\acknowledgments 
We acknowledge the staff of the Accelerator Laboratory of University of Jyv\"askyl\"a (JYFL-ACCLAB) for providing stable online beam. We thank the support by the Academy of Finland with  projects No. 306980, 312544, 275389, 284516, 295207, 314733,  315179, 327629, 320062, 354589, 345869 and 354968.
The support by the EU Horizon 2020 research and innovation program under grant No. 771036 (ERC CoG MAIDEN) is acknowledged.  This project has received funding from the European Union’s Horizon 2020 research and innovation programme under grant agreement No. 861198–LISA–H2020-MSCA-ITN-2019.
V.A.S., O.N., S.S., J.S., and J.K. acknowledge support from the NEPTUN project (PNRR-I8/C9-CF264, Contract No. 760100/23.05.2023 of the Romanian Ministry of Research, Innovation and Digitization).
The work leading to this publication was supported by the Deutsche Forschungsgemeinschaft (DFG, German Research Foundation) - AY 155/2-1.
This project has received funding from the European Union’s Horizon Europe Research and Innovation programme under Grant Agreement No 101057511 (EURO-LABS).

\bibliography{my-final-bib-from-jabref,ref}

\end{document}